\documentstyle[epsf]{mn2e}

\title[Identifying a new IP using XMM-Newton and INTEGRAL]
{Identifying a new intermediate-polar using XMM-Newton and INTEGRAL}

\author[M.Middleton et al.]
{Matthew J. Middleton$^1$, Edward M. Cackett$^2$, Craig Shaw$^1$, Gavin Ramsay$^3$, \newauthor Timothy P. Roberts$^1$ and Peter J. Wheatley$^4$\\
$^1$Department of Physics, University of Durham, South Road, Durham
DH1 3LE,
UK\\
$^2$Institute of Astronomy, University of Cambridge, Madingley Road, Cambridge CB3 0HA,
UK\\
$^3$Armagh Observatory, College Hill, Armagh BT61 9DG\\ 
$^4$Department of Physics, University of Warwick, Coventry CV4 7AL,UK\\
}

\pagerange{\pageref{firstpage}--\pageref{lastpage}} \pubyear{2009}
\long\def\symbolfootnote[#1]#2{\begingroup\def\thefootnote{\fnsymbol{footnote}}\footnote[#1]{#2}\endgroup} 

\begin{document}

\topmargin = -0.5cm

\maketitle

\label{firstpage}

\begin{abstract}

The bright X-ray source, 2XMMi J180438.7-145647 is fortunate to have
long baseline observations in {\it INTEGRAL} that compliment
observations taken by other missions. Optical spectroscopy of this
object has suggested a distance of $\sim$7~kpc and an identification
with a low mass X-ray binary. We instead use the X-ray data from
0.3-40~keV to identify the source as a bright intermediate polar (IP)
with an estimate for the white dwarf mass of
$\sim$0.60~M$_{\odot}$. This identification is supported by the
presence of an iron triplet, the component lines of which are some of
the strongest seen in IPs; and the signature of the spin period of the
white dwarf at $\sim$24 mins. We note that the lack of broad-band
variability may suggest that this object is a stream-fed IP, similar
in many respects to the well studied IP, V2400 Oph. Phase-binning has
allowed us to create spectra corresponding to the peaks and troughs of
the lightcurve from which we determine that the spectra appear harder
in the troughs, consistent with the behaviour of other IPs binned on
their spin periods. This work strongly suggests a mis-identification
in the optical due to the presence of large columns of enshrouding
material. We instead propose a distance to the source of $<$2.5~kpc to
be consistent with the luminosities of other IPs in the dim, hard
state. The relatively high flux of the source together with the
strength of the iron lines may, in future, allow the source to be used
to diagnose the properties of the shock heated plasma and the
reflected component of the emission.

\end{abstract}

\section{Introduction}

Cataclysmic variables (CVs) populate the central-to-faint section of
the galactic X-ray luminosity function (10$^{31-34}$erg s$^{-1}$, see
Sazonov et al. 2006) and may provide a substantial percentage of the
hard X-ray emission seen in the galactic ridge (Revnivtsev et al. 2009). In such systems,
accretion from a low-mass donor star (Cowley et al. 1998 but see also
Orio et al. 2010) onto the white dwarf (WD) may occur via either an
accretion disc (Shakura \& Sunyaev 1973) or via columns of material
following the WD's magnetic field lines and impacting the poles. The
latter system is dubbed a polar or intermediate polar (IP) system
dependent upon the magnetic field strength and the emission is
typified by cooling of shock-heated gas near the WD surface (see
Cropper 1990 for a review of polars) and imprinted absorption and
emission features.

Such systems may be seen to enter a nova outburst state where the
material builds up on the surface of the WD leading to thermo-nuclear
burning (Pietch et al. 2005). The observational result of a nova is a
super-soft source (SSS) but following this they enter a persistent and
much dimmer, harder state (although the long-term behaviour can also
include brighter episodes due to dwarf-novae from disc instabilities,
e.g. Idan et al. 2010). The highest luminosity of this low-state has
been attributed to GK Persei ($\sim$1.1$\times$10$^{34}$ erg s$^{-1}$
in a 2006 outburst) although there is still uncertainty in its
distance ($\sim$ 340~pc - Warner 1987 to 525~pc - Duerbeck 1981) and
this is potentially exceeded by a CV in M3 (1E1339 with a luminosity
of $\sim$1.4$\times$10$^{34}$ erg s$^{-1}$ - Stacey et al. 2011).

\begin{figure*}
\begin{center}
\begin{tabular}{l}
 \epsfxsize=16cm \epsfbox{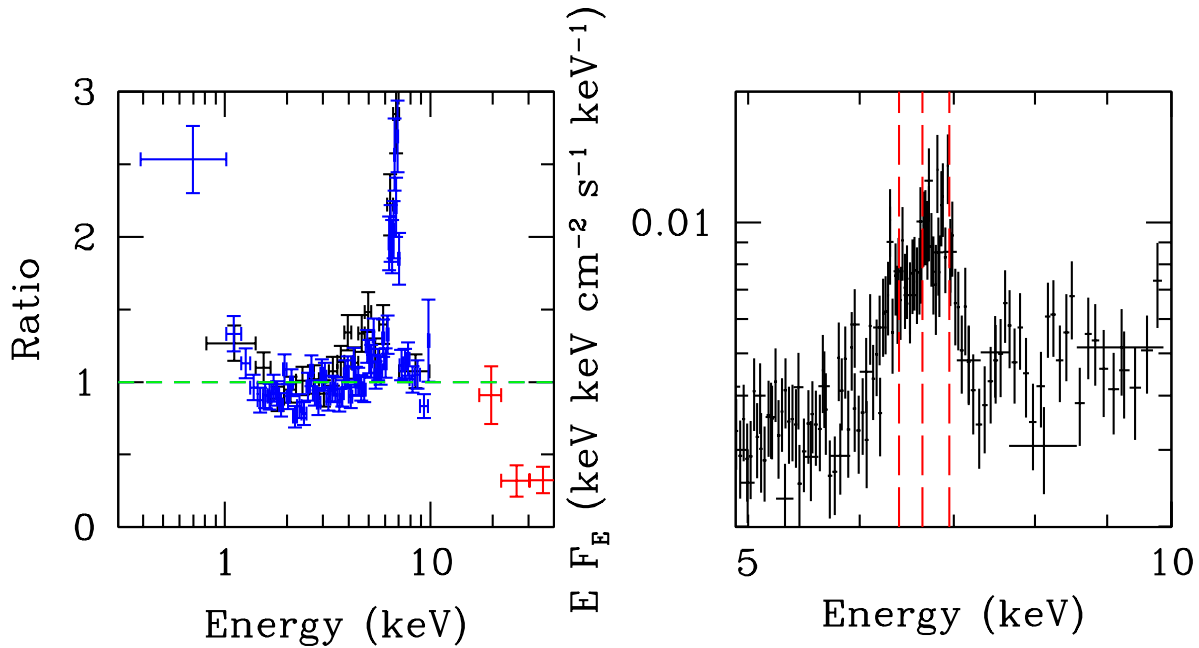}
\end{tabular}
\end{center}
\caption{{\it Left}: ratio of the {\it XMM-Newton} (PN: blue, and MOS1: black)
and {\it INTEGRAL} data (ISGRI: red) to a model of an absorbed
power-law with a best-fitting index of $\sim$0.94. There are clear,
strong residuals due to iron emission and the presence of a roll-over
in the spectrum below $\sim$20~keV. {\it Right}: The {\it XMM-Newton}
data from 5-10~keV (unfolded through a power-law of $\Gamma$=0 and
norm=1) together with dashed vertical lines indicating the rest-frame
energies for the iron triplet (6.4, 6.65 and 6.95 keV respectively).}
\label{fig:l}
\end{figure*}

A further feature of such objects is the presence of periodicity in
the X-ray and optical lightcurves indicating either the orbital period
of the secondary (from the optical emission) ranging from $\sim$1.4 - 50
hours (Hilton et al 2009; Crampton et al. 1986) or the spin period of
the WD, taking values between $\sim$30~s (de Jager et al. 1994) and
$\sim$4000~s (Mauche et al. 2009).

In this letter we present the discovery of a new, persistent, IP using
both {\it INTEGRAL} and {\it XMM-Newton} data to constrain the
important spectral and timing characteristics. We explore the
uncertainty in the distance estimate to this object based on the
optical spectroscopy presented by Masetti et al. (2008) and find that
the estimate of 7~kpc is probably incorrect given the absorbing
material expected to be present in such systems.

\section{Data and previous analysis}

The IR source, 2MASS J180438.92-145647.4 has been identified by the
{\it INTEGRAL} Galactic survey as a hard, X-ray bright source of
interest. Two optical investigations have followed, first by Burenin
et al. (2006) who identify the counterpart as a massive star and
Masetti et al. (2008) who identify it as a low mass star. The latter assume
that the extinction of 3 magnitudes from optical line ratios is
incorrect due to the large inferred distances that would result and
instead determine a distance of $\sim$7 kpc (based on the Galactic
colour excess). This suggests that the source is one of the very
faintest low-mass X-ray binaries (LMXBs) and likely to therefore be a
member of the very-faint X-ray transients (VFXTs see Muno et
al. 2005b; Sakano et al. 2005; Wijnands et al. 2006; Degenaar \&
Wijnands 2009). However, the optical magnitude (18.7 in the R band)
together with 3 magnitudes of absorption would suggest an
identification of the secondary with a giant star and therefore a
high-mass X-ray binary (HMXB). If this is the case then the distance
estimate of 7~kpc would make it extremely faint (only
$\sim$1$\times$10$^{35}$erg s$^{-1}$) for a HMXB (e.g. Grimm, Gilfanov
\& Sunyaev 2006), and would imply that we are looking at an object
even more distant and on the far side of the galactic bulge.

The source has been observed in the 0.2-10~keV X-ray band on three
occasions, a short (1ks) {\it Chandra} positional pointing (OBSID:
7275), a $\sim$12~ks pointed observation by {\it XMM-Newton} (OBSID:
0405390301, identifier: 2XMMi J180438.7-145647) and a followup {\it
Swift} ToO (taken 9th March 2011) which has confirmed that the source
is persistent, thus effectively ruling out an identification with the
subset of VFXTs. The source has also been observed by the higher
energy X-ray detectors on-board {\it INTEGRAL} for a sum exposure time
of $\sim$2500~ks taken over several years overlapping with the {\it
XMM-Newton} observation. Given its proven persistent nature and the
simple fact that the source would not be detected in the higher energy
bandpass if in short-lived outburst (i.e. when a SSS), we include the
data in our analysis as a contemporaneous description of the emission
above 10~keV.

We extract the {\it XMM-Newton} data using {\sc sas v10} and filter
the imaging data using standard patterns ($<$=4 for PN and $<$=12 for
the MOS) and flags (==0). We proceed to remove hard proton flaring
episodes from the full-field high energy ($>$10 keV) lightcurve,
leaving $\sim$ 10ks of good data (this is the most conservative
estimate of the good time interval, the source is actually much
brighter than the background $<$10 keV for the full observation length
of $\sim$12.6~ks), and proceed to extract spectra using {\sc xselect}
from circular source and background regions of 35'' radius. Although
the MOS2 central chip was turned off for the duration of the
observation, both the MOS1 and PN were exposed for the full duration.

\section{X-ray spectral analysis}


\begin{figure*}
\begin{center}
\begin{tabular}{l}
 \epsfxsize=10cm \epsfbox{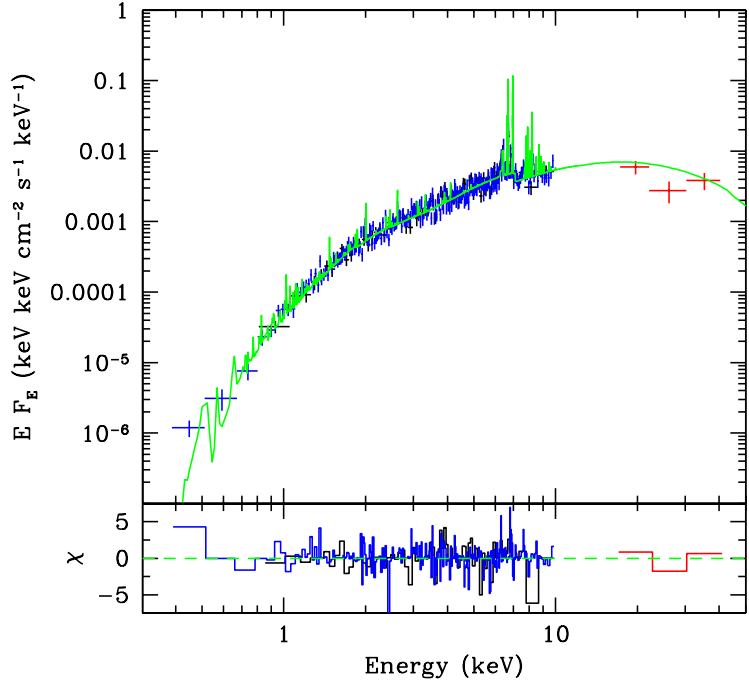}
\end{tabular}
\end{center}
\caption{X-ray spectral data from the PN (blue) and MOS1 (black) together
with the {\it INTEGRAL} ISGRI data (red). The best fitting model
comprising a cooling flow, reflection and absorption (see Table 1) is
shown in green with residuals to this in the panel below.}
\label{fig:l}
\end{figure*}

We fit the background subtracted {\it XMM-Newton} data (grouped on a
minimum of 25 counts per energy channel) together with the standard
spectral {\it INTEGRAL} (ISGRI) products in {\sc xspec v 11.3.2} with
a simple power-law, absorbed by a neutral foreground column and a
constant of proportionality to account for differences between the
detector responses (in {\sc xspec} this is {\sc cons*tbabs*pow}). The
resulting fit quality is very poor ($\chi^2$ of 858.7 for 330 d.o.f.).

The reason for this poor fit is clear from the ratio plot shown in
Figure 1. Whilst the extremely hard ($\Gamma$ = 0.94
$^{+0.05}_{-0.03}$) power-law is a good description of the underlying
continuum below $\sim$6 keV (except at the lowest energies where there
appears to be a soft excess), there is a large, broad iron emission
feature followed by an obvious roll-over below $\sim$20~keV. Adding a
single Gaussian dramatically improves the fit quality ($\Delta\chi^2$
$\sim$ 345 for 3 d.o.f.) giving a broad ($\sigma$ = 0.35~keV) and
strong (equivalent width (EQW) of $\sim$1.4~keV) line which peaks at
6.64~keV. The breadth of the line could be due to smeared reflection
as seen in X-ray binaries (XRBs) at low mass accretion rates (see
Fabian 2005) however, the spectral properties are inconsistent and the EQW of this feature is far larger
than has been seen in any reflection dominated XRB to date, suggesting an alternative origin. Instead we
obtain a marginally better fit ($\Delta\chi^2$ $\sim$12 for 6 d.o.f.)
by describing the excess as an iron triplet corresponding to lab-frame
energies of $\sim$6.4, 6.65 and 6.95 keV. Together with the underlying
hard continuum ($\Gamma\sim$1.00$\pm$0.05), we then initially identify the
source as an IP in the persistent dim, hard state (rather than a polar
candidate as the spectrum is too hard to be described by cyclotron
emission: Cropper 1990). In this case we would expect the spectrum to
be dominated by Bremsstrahlung cooling with a strong component of
photo-electric absorption (Norton \& Watson 1989) and reflected
emission from the WD surface creating the strong iron K$_{\alpha}$
line (Beardmore et al. 1995; Done, Osborne \& Beardmore 1995; Done \&
Magdziarz 1998).

Although a good physical description is one of a multi-temperature
model for the plasma, generally speaking the data quality of IPs is
not high enough to distinguish this from a single temperature plasma
model. We therefore assume a simple, single temperature description of
the plasma flow, and create a model comprising a cooling flow plasma,
reflection from the surface of the WD (with a fixed incident spectral
index of 1.5) all of which is obscured by a neutral partial covering
fraction and a foreground Galactic column\footnotemark\footnotetext{We
use an extrinsic column with an upper limit of
0.6$\times$10$^{22}$cm$^{-2}$ to be consistent with the estimates
available from HEASARC (Dickey \& Lockman 1990; Kalberla et al. 2005)}
(in {\sc xspec} this is {\sc
cons*tbabs*pcfabs*(refxion\symbolfootnote[2]{This is a private model
based on {\sc reflbal} as described in Done \& Gierli{\'n}ski (2006)
but with the updated ionisation tables of Ross \& Fabian
(2005).}(cvmekal\symbolfootnote[3]{See Singh et al.  (1996).}))}). We
find the requirement for a second neutral covering fraction
($\Delta\chi^2$ $\geq$100 for 2 d.o.f.) although the inclusion of any further
absorbers does not significantly improve the fit. We obtain an good
description of the data ($\chi^2$ of 346.5 for 325 d.o.f.)  with the
best-fitting model and residuals plotted together in Figure 2 and the
model parameters with 90\% confidence limits provided in Table 1 (note
that, in this best-fitting case, the constants of proportionality are
consistent within 10\% of unity). The model provides an estimate for
the unabsorbed 0.3-10 keV flux of 2.4$\times$10$^{-11}$ ergs cm$^{-2}$
s$^{-1}$. This is very bright for an IP in the dim, hard state
(although a distance of $\sim$2.5 kpc would be required in order for
this to be the brightest: Stacey et al. 2011). However, whilst the
source is relatively bright, the hard continuum precludes a useful RGS
analysis which could identify soft emission/absorption features (and
identify the multi-temperature nature of the
emission: Wu et al. 2003).

The strong iron feature in the spectrum (see Fig. 1) may be well
described by a combination of 3 Gaussians (although the 6.4~keV line
originates from reflection from the WD surface rather than intrinsic
photo-electric emission from the plasma) and we obtain the properties
of these by fitting the {\it XMM-Newton} data alone together with a
simple absorbed power-law continuum. The best-fitting model parameters
are also shown in Table 1. These show that, whilst the properties of
the continuum are consistent with those seen for other IPs, the
inferred equivalent widths (EQW) of the iron lines are towards the
upper limit of what has previously been observed (e.g. Butters et
al. 2011). We note that, due to the marginal improvement in fit quality from
including 3 Gaussians over a single component, obtaining constraining
estimates for the errors on the strength (EQW) and breadth ($\sigma$)
is unrealistic and so are omitted.

To ensure that cross-calibration errors are not overly influencing the
best-fitting model parameters, we test the upper and lower limits for
the constant of proportionality to the {\it INTEGRAL} data
($\sim$1.03$\pm$0.4 - consistent with the cross-calibration status of
EPIC versus ISGRI\footnotemark\footnotetext{See Kirsch et al. (2004) and
http://heasarc.nasa.gov/docs/heasarc/caldb/caldb\_xcal.html. For examples of this in use see e.g. Caballero-Garc{\'{i}}a et al. (2009)}. In both
cases, we obtain values for the temperature of the rollover within the
90\% confidence limits presented in Table 1. This rollover in the
spectrum, due to the mean temperature of the plasma emission, can then
be used as a crude measure for the mass of the WD. Aizu (1973) gives a
formula relating the gravitational potential of the WD to the mass and
radius:

\begin{equation} 
{\rm kT}_{\rm s} = 16\times\left(\frac{M}{0.5 M_{\odot}}\right)\left(\frac{R}{10^{9}{\rm cm}}\right)^{-1} {\rm keV}
\end{equation} 

\noindent Substituting the analytical expression for the radius (Nauenberg 1972):
\begin{equation} 
{\rm R} = 0.78\times10^{9}\left[\left(\frac{1.44 M_{\odot}}{M}\right)^{\frac{2}{3}} -
\left(\frac{M}{1.44 M_{\odot}}\right)^{\frac{2}{3}}\right]^{\frac{1}{2}} {\rm cm}
\end{equation} 

\noindent allows us to calculate the mass of the WD from the shock
temperature:

\begin{equation} 
\frac{M_{\rm WD}}{M_{\odot}} = 1.44\times\left[\frac{1}{2}\left(1+\sqrt{1+4\times\left(\frac{59}{KT_{\rm s}}\right)^{2}}\right)\right]^{-\frac{3}{4}}
\end{equation}

\noindent Taking the best-fitting value for kT$_{\rm s}$ from our
model (23.30 $_{-5.09}^{+8.84}$ keV), this gives a mass of
0.62$_{-0.09}^{+0.13}$M$_{\odot}$ assuming a single temperature
description of the cooling flow. Based on the correction factor expected from the change in the gravitational potential over the shock height (Cropper et al. 1998, 1999) this provides an estimate for the mass of $\sim$0.60~M$_{\odot}$, consistent with the median mass of isolated WDs (Kepler et al. 2007). It should be noted however that this assumes a single-temperature description whereas a physical model may well place the peak shock temperature above this (Cropper et al. 1999).

An important consequence as this identification is that there is a
clear prediction for the presence of large columns of material that
obscure the intrinsic emission (see Cropper et al. 1990). This
therefore makes any associated optical identification and distance
estimate based on line spectra highly unreliable as the intervening
material will distort the emission. Instead, in order for the source
to have similar luminosities to those of other IPs in the dim, hard
state, we expect the distance to be $<$2.5~kpc.

\begin{table}
\begin{center}
\begin{minipage}{80mm}
\bigskip
\caption{Best fitting spectral parameters}

\begin{tabular}{l|c}
  \hline


\multicolumn{2}{|c|}{{\sc cons*tbabs$_1$*pcfabs$_1$*pcfabs$_2$(refxion(cvmekal))}}\\
\hline
n$_{\rm H,1}$ ($\times$10$^{22}$cm$^{-2}$) & 28.02  $_{-6.88}^{+8.51}$\\
Fraction$_1$  & 0.67 $_{-0.06}^{+0.05}$\\
n$_{\rm H,2}$ ($\times$10$^{22}$cm$^{-2}$) &  1.85 $_{-0.40}^{+0.42}$\\
Fraction$_2$  & 0.80 $_{-0.03}^{+0.06}$\\
log$\xi$ & 1.00 $_{-peg}^{+0.77}$\\
kT$_{max}$ (keV) & 23.30 $_{-5.09}^{+8.84}$\\
Fe abund& 0.67 $_{-0.17}^{+0.25}$\\
& \\

$\chi^{2}$ (d.o.f.) & 346.8 (324)\\
Null P & 0.18 \\
\hline
\multicolumn{2}{|c|}{{\sc cons*tbabs(powerlaw+gauss$_1$+gauss$_2$+gauss$_3$)}}\\
\hline
n$_{\rm H,1}$ ($\times$10$^{22}$cm$^{-2}$) & 0.81$_{-0.07}^{+0.06}$ \\
Line energy$_1$ (keV)& 6.41 $\pm$ 0.06\\
EQW$_1$ (eV)& 244\\
$\sigma_1$ (keV)& 0.16\\ 
Line energy$_2$ (keV)& 6.69 $\pm$ 0.04\\
EQW$_2$ (eV)& 99\\
$\sigma_2$ (keV)& 3.8$\times$10$^{-4}$\\ 
Line energy$_3$ (keV)& 6.91 $\pm$ 0.05\\
EQW$_3$ (eV)& 324 \\
$\sigma_3$ (keV)& 7.2$\times$10$^{-2}$\\ 
$\Gamma$ & 0.67 $\pm$ 0.05\\
norm & 3.2$\times$10$^{-4}\pm$0.2$\times$10$^{-4}$\\

$\chi^{2}$ (d.o.f.) & 353.7 (316)\\
Null P &  0.07 \\

  \hline

\end{tabular}
Notes: The best-fitting parameter values are given with 90\%
confidence limits with the properties of the iron lines obtained
separately to those of the continuum. Note that the ionisation parameter (log$\xi$) is set to peg at a lower limit of 1.

\end{minipage}
\end{center}
\end{table}

\section{Timing analysis}

There is a wealth of evidence suggesting the presence of periodicities
from the optical and X-ray lightcurves of IPs, attributed to either
the orbital period of the secondary or the spin period of the
WD\footnotemark\footnotetext{http://asd.gsfc.nasa.gov/Koji.Mukai/iphome/catalog/members.html}. The
properties of the broad-band noise on which these periodicities sit in
the power density spectrum (PDS) has been of great interest as the
behaviour can provide observational tests for theories of how
non-linear variability is generated in the accretion process (see
e.g. the discussion of Uttley, McHardy \& Vaughan 2005). In the case of XRBs, this
is now widely accepted to be due to propagating fluctuations in the
disc (see Lyubarskii 1997, Ar{\'e}valo \& Uttley 2006, Ingram \& Done
2011). However, in the case of IPs, the inner regions of the inflow
are disrupted by the magnetic field of the WD, leading to a large
truncation radius of the accretion disc, within which, the material
follows the magnetic field lines. There is therefore the prediction of
a break in the PDS (beyond which the red noise follows $\sim\nu^{-2}$:
Revnivtsev et al. 2010) which has now been observationally confirmed
(Revnivtsev et al. 2009, 2010). Due to this break at low frequencies
there is very little rapid variability, unlike in the cases of XRBs
where significant rapid variability can be created by the
magneto-rotational instability (MRI, see Beckwith, Armitage \& Simon 2011) being established in a low
density flow (Done, Gierli{\'n}ski \& Kubota 2007). In the cases where
the accretion flow is stream-fed (e.g. V2400 Oph), the lack of
broad-band noise is evidently due to the observational lack of an
accretion disc (Hellier \& Beardmore 2002). The position of the
truncation radius and the nature of the accretion are therefore
paramount in accurately determining the presence of periodicities. For
example, where the disc truncates at very large radii
(i.e. effectively stream-fed) there is the prediction that, at high
frequencies, there should be little or no red noise in the PDS making
significance tests here rather ambiguous (see the discussion of
Vaughan 2005). However, where the statistical quality of the data
allows, and there is a clear (and stringently tested) lack of
broad-band red noise, there may be an argument for a significance test
to be carried out above the white noise. In such a situation,
observing the behaviour of the spectra binned on the period may allow
for a second order test (see following section).

Although we do not have any simultaneous optical data, the
identification of an X-ray periodicity may help further constrain the
identification with an IP. We extract the 0.3-10 keV, co-added
(using the PN and MOS1 with the same start and stop times), background
subtracted lightcurve binned on 10~s, using the same regions as the
spectral analysis. As the source dominates the background emission
throughout the observation, we use the full observation length in
order to improve any variability constraints. We initially test the
shape of the broad-band variability out to the longest available
timescales using the most robust LS method available (Vaughan 2010)
and find that the broad-band continuum is consistent with the white
noise at the 3$\sigma$ level (although we identify a possible peak of
excess variance at $\sim$ 0.7~mHz). Although we cannot probe down to
very low frequencies, the fairly high data quality from this
observation would imply that we should see variability power on these
frequencies if there was a contribution from disc
accretion (Revnivtsev et al. 2010).  This suggests that we are
observing a stream-fed IP similar to V2400 Oph (Norton, Haswell, \&
Wynn 2004) which is seen to possess a similar broad-band X-ray
spectrum (Revnivtsev et al. 2004).  This lack of broad-band noise
therefore allows us to obtain tentative significance measurements
assuming a white noise description of the continuum variability. We
proceed to extract the power density spectrum (PDS) using the {\sc
ftool: powspec} in units of fractional rms$^{2}$. We extract the PDS
averaged over 5 segments of the lightcurve in order to realistically
constrain any bins of variability. We again find an excess of variance
at $\sim$0.7~mHz with the tightest constraints in the 0.8-5~keV band
at a significance $>$3$\sigma$ (see Fig. 3). We note that the
coherence of this excess variance ($\nu_{QPO}/\nu_{FWHM}$) is rather
low, preventing a robust identification with a period/quasi-period
(although the behaviour of the broad-band noise is unlikely to
facilitate an identification with a break in the PDS), however a
$\sim$24 min period could be readily associated with the spin period
of the WD and is similar to that of V2306 Cyg (Norton et al. 2002) at
similar flux levels.

\begin{figure}
\begin{center}
\begin{tabular}{l}
 \epsfxsize=8cm \epsfbox{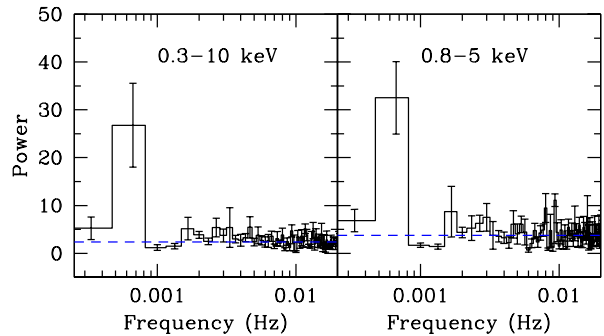}
\end{tabular}
\end{center}
\caption{Average PDS extracted from the 0.3-10~keV (left) and
  0.8-5~keV (right) lightcurves binned on 10~s, made from 5 segments
  of the lightcurve including the white noise (horizontal dashed
  line). There is an excess of variance at $\sim$0.7~mHz which is
  significant at $>$3$\sigma$ in the 0.8-5~keV band. As the power over
  the remaining observable frequency bins is consistent with the
  statistical white noise (possibly indicating the source as being
  stream-fed) this is an acceptable method for determining the
  significance of this feature.}
\label{fig:l}
\end{figure}

\subsection{Phase-binned spectroscopy}

We can better confirm the nature of the excess variance by extracting
the spectral properties in the peaks and troughs of the
lightcurve. This phase-binned spectroscopy has been applied to a
number of IPs with confirmed spin periods (e.g Pek{\"o}n \& Balman 2010; Hellier et al. 1996; Done, Osborne \& Beardmore 1995) and, in each case,
the troughs are found to have harder spectra due to the changing
proportions of reflection and absorption.

We follow a similar approach and extract PN spectra from epochs
corresponding to the peaks and troughs ensuring no overlap between the
two and avoiding the flaring episodes as with the time-averaged
spectroscopy. This gives a total exposure of 3.8~ks for the peaks and
5.7~ks for the troughs. To identify any changes in continuum shape, we
apply a simple power-law model to both sets of data simultaneously
({\sc (tbabs(powerlaw)}) over the 0.3-6~keV band (to ignore the
distorting effects of the iron emission). We obtain best-fitting
spectral index values of 0.32$_{-0.12}^{+0.12}$ and
0.57$_{-0.14}^{+0.15}$ for the troughs and peaks respectively and
an improvement in $\Delta\chi^2$ of $\sim$4.8 for 1 d.o.f. over fixing
the index to be the same in both. This allows us to claim that the
spectrum of the troughs appears harder, consistent with the
behaviour of other IPs in the dipping phase of their spin period. We
attempt to quantify the changing properties of the iron emission by
fixing the continuum to the best-fitting value below 6~keV and
including 3 Gaussian components in the model. However, the much
shorter exposure for each phase-binned spectrum prevents any
difference from being well constrained.

\section{Conclusion}

The hard, bright X-ray source, 2XMMi J180438.7-145647, has been
observed only once by {\it XMM-Newton}, however, together with the
available {\it INTEGRAL} (ISGRI) data, this relatively short observation has
shown that the time-averaged X-ray emission is characteristic of an
IP. In general we determine similar properties for this source when we
compare them with that of the wider population of IPs, save for the iron features which are notably strong. It is
possible that any future, longer observation, may be able to use this
strong iron triplet to provide a better diagnostic of the plasma
region.

Due to the extremely hard X-ray spectrum and short observation length,
we cannot obtain a useful RGS spectrum thus precluding a more complex
(multi-temperature) description of the plasma. However, the inclusion
of {\it INTEGRAL} (ISGRI) data has allowed us to place constraints on the 
mean temperature of the plasma cooling flow. Taking into account the
changing gravitational potential over the shock height, this provides a
an estimate for the mass of the WD of $\sim$0.60~M$_{\odot}$,
consistent with the median of isolated WD masses.

We detect the presence of excess variance power at $\sim$0.7~mHz which
{\it may} be associated with the spin period of the WD. Phase-binning
the emission produces spectra which show behaviour consistent with
that of other IPs, supporting this identification for the source. The
lack of variability apart from this feature suggests that we are
observing a stream-fed IP, similar in behaviour to V2400 Oph.

Given the large columns of absorbing material enshrouding IPs, we find
that the distance estimate based on the
optical spectrum (Masetti et al. 2008) to be highly
unreliable. Instead, given the flux of the source, its distance is
much more likely to be $<$2.5~kpc to be consistent with the observed
luminosities of other IPs.

\section{Acknowledgements}

MM and TR thanks STFC for support in the form of a standard grant. This work
is based on observations obtained with {\it XMM-Newton}, an ESA
science mission with instruments and contributions directly funded by
ESA Member States and NASA. This work also makes use of observations
taken by {\it INTEGRAL}, an ESA project with instruments and science
data centre funded by ESA member states (especially the PI countries:
Denmark, France, Germany, Italy, Switzerland, Spain), Poland and with
the participation of Russia and the USA. We also acknowledge support from Dr Tony Bird at the University of Southampton for his help with the {\it INTEGRAL} data products.

\label{lastpage}

\end{document}